\documentclass[letter,10pt,twocolumn]{article}
\usepackage{float}
\usepackage{graphicx}
\usepackage{dcolumn}
\usepackage{amsmath}
\usepackage[dvips]{epsfig}
\usepackage{subfigure}

\begin{document}

\title{Optimization of three-dimensional micropost microcavities for cavity quantum electrodynamics}
\author{Jelena Vu\v{c}kovi\'{c}, Matthew Pelton, Axel Scherer, and Yoshihisa Yamamoto\\
Quantum Entanglement Project, ICORP, JST\\
         Edward L. Ginzton Laboratory, Stanford University\\
         Stanford, California, 94305 U.S.A.}

\maketitle

\begin{abstract}
\vbox{\vspace{0.1in}} This article presents a detailed analysis,
based on the first-principles finite-difference time-domain
method, of the resonant frequency, quality factor ($Q$), mode
volume ($V$), and radiation pattern of the fundamental ($HE_{11}$)
mode in a three-dimensional distributed-Bragg-reflector (DBR)
micropost microcavity. By treating this structure as a
one-dimensional cylindrical photonic crystal containing a single
defect, we are able to push the limits of $Q/V$ beyond those
achievable by standard micropost designs, based on the simple
rules established for planar DBR microcavities. We show that some
of the rules that work well for designing large-diameter
microposts ({\it e.g.}, high-refractive index contrast) fail to
provide high-quality cavities with small diameters. By tuning the
thicknesses of mirror layers and the spacer, the number of mirror
pairs, the refractive indices of high and low refractive index
regions, and the cavity diameter, we are able to achieve $Q$ as
high as $10^4$, together with a mode volume of $1.6$ cubic
wavelengths of light in the high-refractive-index material. The
combination of high $Q$ and small $V$ makes these structures
promising candidates for the observation of such cavity quantum
electrodynamics phenomena as strong coupling between a quantum dot
and the cavity field, and single-quantum-dot lasing.
\end{abstract}


\section{Introduction}
Spontaneous emission is not an intrinsic property of an isolated
atom, but is rather a property of an atom coupled to its
electromagnetic vacuum environment. The spontaneous emission rate
is directly proportional to the density of electromagnetic states
that a spontaneously emitted photon can couple to, and can be
modified with respect to its value in free space by placing the
atom in a cavity \cite{ref:Purcell}. The experimental
demonstrations of the inhibition and enhancement of spontaneous
emission rate were carried out starting in the mid-70's
\cite{ref:Drexhage74,ref:Kleppner81,ref:Paris83,ref:Gabrielse85,ref:Rome87,
ref:MIT87}, using atoms coupled to single mirrors, planar
cavities, or spherical Fabry-Perot resonators. Advances in
microfabrication techniques enabled the construction of
high-quality semiconductor micropost and microdisk microcavities
in the late 80's and early 90's, and ignited interest in
solid-state cavity quantum electrodynamics (QED) experiments
\cite{ref:Yamamoto89,ref:VCSEL,ref:microdisk}. In 1987,
photonic-crystal structures were proposed as promising candidates
for strong spontaneous emission modification
\cite{ref:Yablonovitch87, ref:SJohn87}, but the first experimental
results on photonic crystal microcavities were not reported until
a decade later \cite{ref:Baba97,ref:Foresi97}.

Cavity-QED phenomena in the low-$Q$ (weak-coupling) regime, as
well as in the high-$Q$ (strong-coupling) regime, can be used in
construction of high-efficiency light-emitting diodes,
low-threshold lasers, and single photon sources. A powerful
property of solid-state microcavities is that a single
narrow-linewidth emitter (quantum dot) can be embedded in them
during the growth process, enabling cavity-field interaction with
such artificial atom \cite{ref:Solomon01}. Due to imperfections in
fabricated structures, un-optimized structure parameters, and the
inability to precisely control position of a quantum dot, only
phenomena in the low-$Q$ regime have been observed so far.

The first successful optical characterizations of photonic-crystal
microcavities with quantum dots were performed recently
\cite{ref:CJMSmith99,ref:PBGqdot_UCSB,ref:PBGqdot_Caltech1}. $Q$
factors as large as $2800$ were reported, together with mode
volumes as small as $0.5(\lambda/n)^3$, where $\lambda$ is the
optical wavelength, and $n$ is the refractive index of the
dielectric material \cite{ref:PBGqdot_Caltech2}. The possibility
of improving the quality factor while preserving such a small mode
volume makes these structures good candidates for cavity QED, in
particular with neutral atoms (due to a strong field intensity in
the air region for the optimized cavity designs)
\cite{ref:JV2001,ref:JV2002}. So far, this has not been
demonstrated experimentally.

\begin{figure}[htbp]
\begin{center}
\epsfig{figure=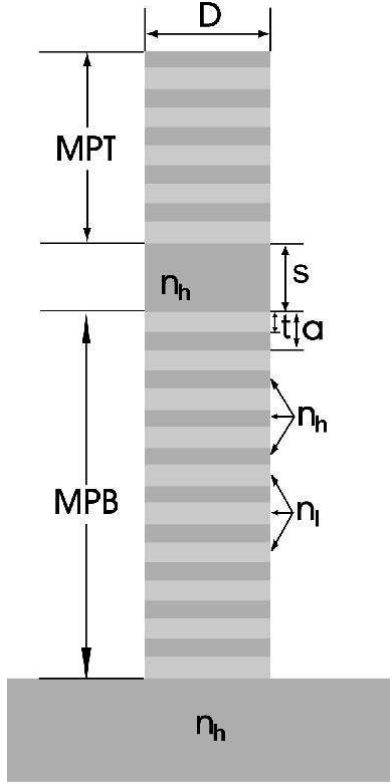, width=2in} \caption{Parameters for a micropost microcavity. The microposts analyzed in this paper are rotationally symmetric
around the vertical axis.} \label{fig:structure}
\end{center}
\end{figure}

The advantages of microposts relative to other microcavities are
that the light escapes in the normal direction to the sample in a
single-lobed Gaussian-like pattern, and that it is relatively
straightforward to isolate a single quantum dot in a post.
However, in order to observe such cavity QED phenomena as strong
coupling with a single dot or single-dot lasing in these
structures, a number of design and fabrication issues have to be
addressed. In this article, we present the optimization of
micropost parameters (illustrated in Fig. \ref{fig:structure}), in
order to maximize the quality factor and minimize the volume of
the fundamental ($HE_{11}$) mode (whose field pattern is shown in
Fig. \ref{fig:EBfield}). We show that both strong coupling cavity
QED with a single quantum dot, and single-quantum-dot lasing are
possible in the optimized micropost microcavity.

All analyzes presented in this article are performed by the
finite-difference time-domain (FDTD) method, which enables
accurate modelling of the electromagnetic properties of structures
with complex geometries. The rotational symmetry of micropost
microcavities allows us to use a cylindrical FDTD algorithm and
reduce the order of the computer memory requirements from $N^3$ to
$N^2$, where $N$ represents a linear dimension of the
computational domain. The method used is described in detail in
our earlier publication \cite{ref:Pelton02}.

\section{Motivation for maximizing the ratio of quality factor to mode volume}

Let us assume that a single quantum dot is isolated in a
microcavity, and that the transition frequency from the
one-exciton state to the zero-exciton state is on resonance with
the fundamental optical cavity mode frequency $\omega$. Under
these conditions, the system can be modelled in the same way as a
single two-level atom coupled to a single cavity mode, and
described by the Jaynes-Cummings Hamiltonian
\cite{ref:Yamamoto_Imamoglu}. The coupling parameter $g$ between
the exciton and the cavity field reaches its maximum value equal
to the vacuum Rabi frequency $g_0$, when the dot is located at the
point of the maximum electric field intensity, and when the
excitonic dipole moment is aligned with the electric field:
\begin{equation}
g_0=\frac{\mu}{\hbar} \sqrt{\frac{\hbar\omega}{2\epsilon_M V}},
\label{eqn:g}
\end{equation}
where $\epsilon_M$ is the dielectric constant at the location of
the exciton, $\mu$ is the dipole moment matrix element between the
one-exciton and zero-exciton states, and $V$ is the cavity mode
volume, defined as
\begin{equation}
V=\frac{\iiint\epsilon(\vec{r})|E(\vec{r})|^2d^3\vec{r}}{
\max\bigl [  \epsilon(\vec{r})|E(\vec{r})|^2 \bigr ]}.
\end{equation}

Depending on the ratio of the coupling parameter $g$ to the {\it
cavity field decay rate} $\kappa=\omega /2Q$ and the {\it
excitonic dipole decay rate} $\gamma$, we can distinguish two
regimes of coupling between the exciton and the cavity field: {\it
strong coupling}, for $g>\kappa,\gamma$, and {\it weak coupling},
for $g<\kappa,\gamma$. In the strong-coupling case, the exciton is
coherently coupled to the cavity field, spontaneous emission is
reversible, and {\it vacuum Rabi oscillation} occurs. On the other
hand, in the weak-coupling case, the spontaneous emission is
irreversible, and the spontaneous emission decay rate $\Gamma$ is
\cite{ref:Yamamoto_Imamoglu}
\begin{equation}
\Gamma=g^2\frac{4Q}{\omega}.
\end{equation}

The spontaneous emission rate of an exciton in free space, on the
other hand, is given by
\begin{equation}
\Gamma_0=\frac{\omega^3\mu^2}{3\pi\epsilon_0\hbar c^3} \, .
\label{eqn:gamma0}
\end{equation}
The ratio of $\Gamma$ to $\Gamma_0$ is called the {\it Purcell
factor} \cite{ref:Purcell}. For an exciton positioned at the
maximum of the field intensity and aligned with the electric
field, the Purcell factor is equal to
\begin{equation}
F_0=\frac{3Q\lambda^3\epsilon_0}{4\pi^2V\epsilon_M}.
\label{Purcell}
\end{equation}

 We usually define the Purcell factor $F$
as the spontaneous emission rate enhancement relative to the bulk
material. The spontaneous emission rate in the bulk material with
refractive index $n_h$ is enhanced $n_h$ times with respect to its
value in free space, which implies that $F=F_0/n_h$.

If the Purcell factor is much greater than one, the exciton will
radiate much faster in the cavity than in free space. The
radiative-rate enhancement is proportional to the ratio of the
quality factor to the volume of the cavity mode, according to Eqn.
\ref{Purcell}. The Purcell factor increases with $Q/V$ only to the
point where the coupling parameter $g$ becomes larger than the
decay rates of the system ($\kappa$ and $\gamma$). At that point,
the coupled exciton-cavity system enters the strong-coupling
regime.

\begin{figure}[hbtp]
\begin{center}
\subfigure{\epsfig{figure=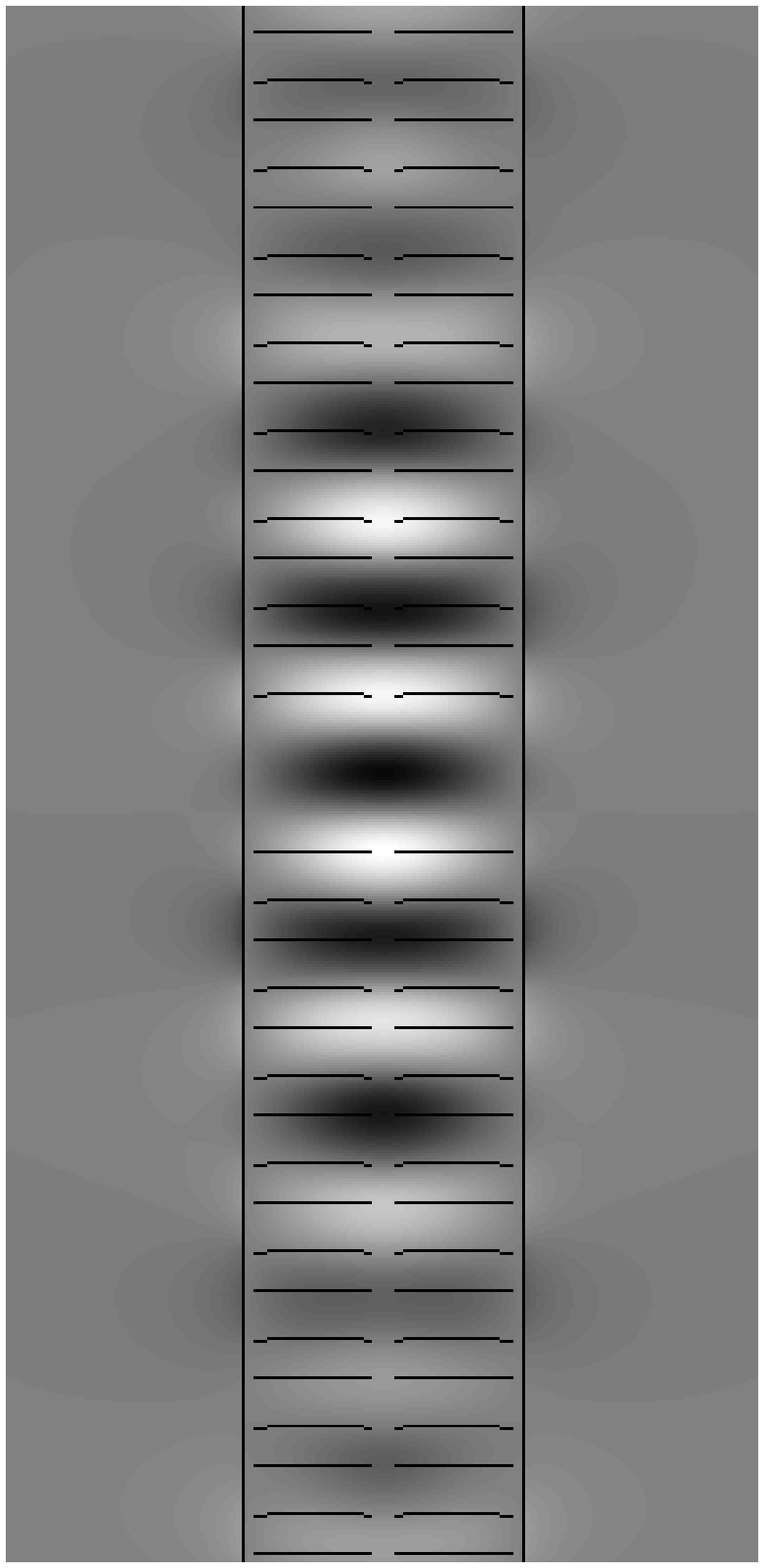, width=1.5in}} \subfigure{\epsfig{figure=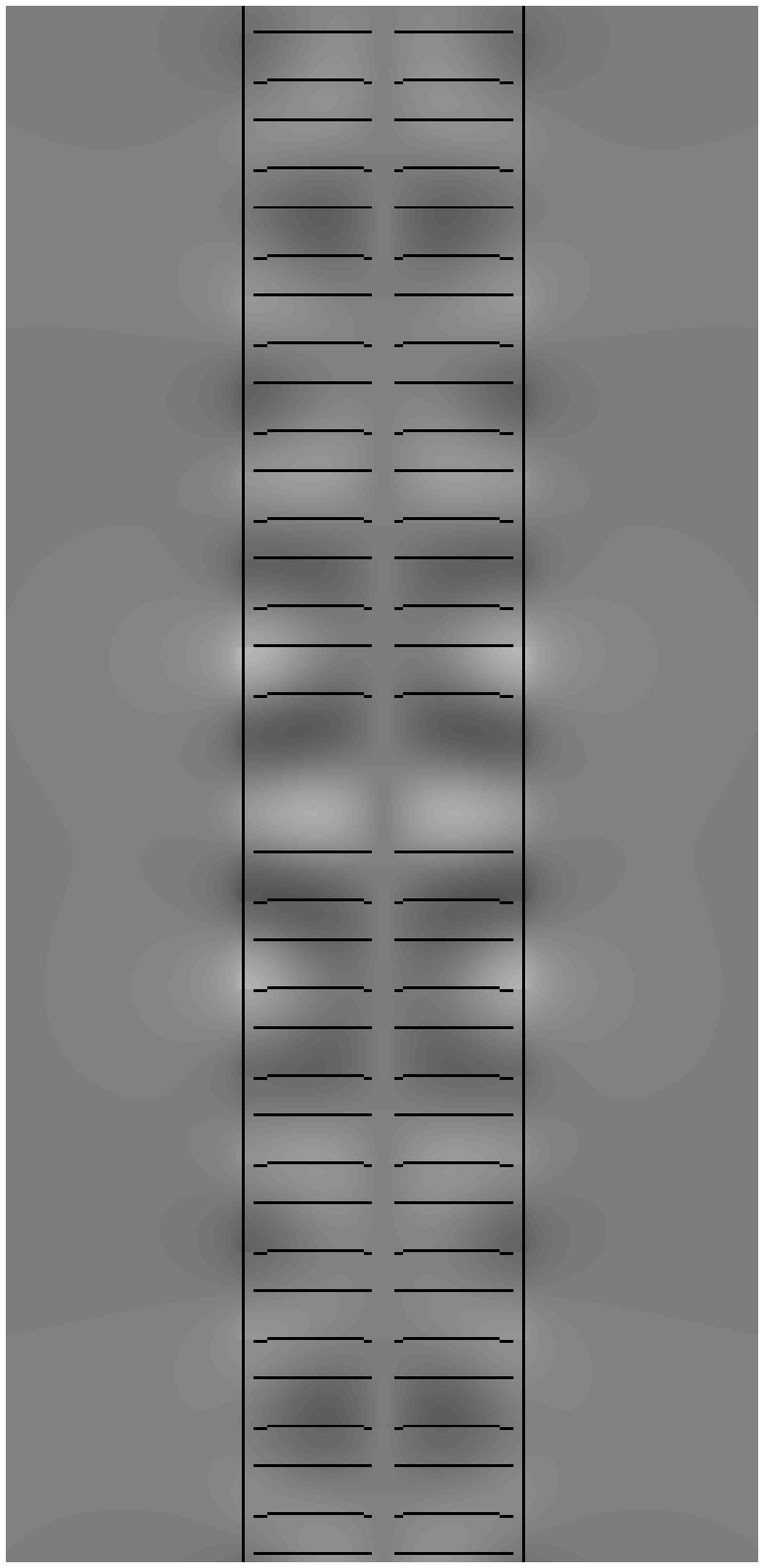, width=1.5in}} \vspace{0.3cm}
\subfigure{\epsfig{figure=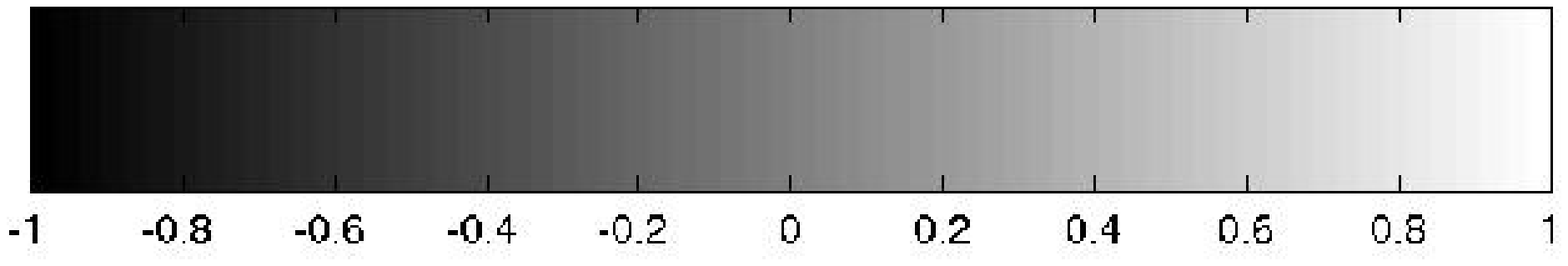, width=3in}}\caption{Electric-field components for the fundamental ($HE_{11}$) mode in a micropost microcavity. The left
figure illustrates the electric-field component parallel to the distributed Bragg reflectors (DBR's), while the figure on the right represents the electric field
component perpendicular to the DBR's. The micropost parameters, using the notation from Fig. \ref{fig:structure}, are as follows: the cavity diameter $D=0.5
\mu$m, the refractive indices of high/low refractive index regions $n_h=3.57$ / $n_l=2.94$, the DBR periodicity $a=155$ nm, the thickness of the low refractive
index mirror layer $t=85$ nm, the spacer thickness $s=280$ nm, and the number of mirror pairs on top/bottom $MPT=15$ / $MPB=30$.} \label{fig:EBfield}
\end{center}
\end{figure}

Increasing $Q/V$ can also lead to a reduction in laser threshold.
The fraction of the light emitted by an exciton that is coupled
into one particular cavity mode is known as the spontaneous
emission coupling factor $\beta$, and is related to the Purcell
factor via the following expression:
\begin{equation}
\beta=\frac{F}{1+F}.
\end{equation}
Therefore, if the emission rate of an exciton is strongly enhanced
by its interaction with a cavity mode, the fraction of spontaneous
emission going into all other modes ($1-\beta$) is reduced. The
fraction of spontaneous emission going into non-lasing modes is
one of the fundamental losses in a laser, and by decreasing it,
one can lower the laser threshold.

Of particular interest would be a single-dot laser, which
represents an ultimate microscopic limit for semiconductor lasers.
The realization of such a device would allow physical
investigations similar to those afforded by the single-atom laser
\cite{ref:salaser}. Lasing of such a microscopic system would
occur when the mean spontaneously-emitted photon number $n_{sp}$
in the laser mode becomes larger than one \cite{ref:Bjork94}:
\begin{equation}
n_{sp}=\frac{\beta\tau_{ph}N_A}{\tau_{sp}}=\frac{N_A(\Gamma_0
n_h)}{ \omega / Q }\cdot\frac{F^2}{1+F} \geq 1,
\label{eqn:threshold}
\end{equation}
where $\tau_{sp}=1/\Gamma$, $\tau_{ph}=Q/\omega$, and $N_A$ is the
average probability over time that the quantum dot contains an
exciton.

One of the most interesting applications of cavity QED is the
construction of efficient sources of single photons
\cite{ref:Santori01,ref:Michler00,ref:Gerard01}. Single-photon
sources are useful for quantum cryptography \cite{ref:BB84},
quantum computation \cite{ref:Yamamoto88,ref:qucomp_prop1},
quantum networking \cite{ref:qucomp_prop2}, and random number
generators {\cite{ref:randnos1, ref:randnos2}. A single quantum
dot can be used to generate single photons, and the output
coupling efficiency can be enhanced by cavity QED. In other words,
by changing the cavity parameters ($Q/V$) and the quantum dot
location, we can control the probability of coupling this
spontaneously emitted single photon into the mode of interest, and
subsequently coupling it into the communication channel.

\section{Micropost microcavities}

Micropost microcavities consist of a high-refractive-index region
(spacer) sandwiched between two dielectric mirrors, as shown in
Fig. \ref{fig:structure}. Confinement of light in these structures
is achieved by the combined action of distributed Bragg reflection
(DBR) in the longitudinal direction (along the post axis), and
total internal reflection (TIR) in the transverse direction (along
the post cross-section). The microposts analyzed in this paper are
rotationally symmetric around the vertical axis. The DBR mirrors
can be viewed as one dimensional (1D) photonic crystals generated
by stacking high- and low-refractive-index disks on top of each
other. The microcavity is formed by introducing a defect into this
periodic structure. The periodicity of the photonic crystal is
denoted as $a$, the thickness of the low-refractive-index disks is
$t$, the diameter of the disks is $D$, and the refractive indices
of the low- and high-refractive-index regions are $n_l$ and $n_h$,
respectively. The defect is formed by increasing the thickness of
a single high-refractive-index disk from $(a-t)$ to $s$, as shown
in Fig. \ref{fig:structure}. The number of photonic crystal
periods above and below the defect region ({\em i.e.}, the number
of DBR pairs) is labelled as $MPT$ and $MPB$, respectively.

The mode of interest to us is the doubly-degenerate fundamental
($HE_{11}$) mode, whose field pattern is shown in Fig.
\ref{fig:EBfield}. The parallel component of the electric field is
dominant in this mode, and has an antinode in the center of the
spacer. Furthermore, in this central plane, the electric field is
practically linearly polarized along the vertical axis of the
micropost, while there is a small deviation from the linear
polarization at larger distances from this axis.

\begin{figure}[hbtp]
\begin{center}
\epsfig{figure=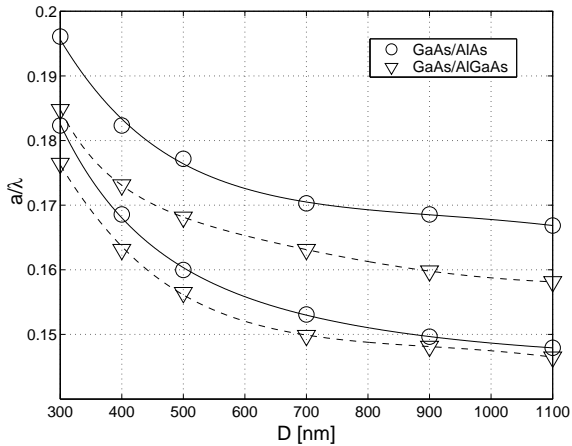, width=3in}\caption{Band gap edges, calculated using the FDTD method (points), of the fundamental ($HE_{11}$) mode in a
cylindrical one-dimensional photonic crystal in the GaAs/AlAs or GaAs/Al$_{x}$Ga$_{1-x}$As material systems. The lines are guides to the eye. The GaAs/AlAs
photonic crystal has the following parameters: $n_h=3.57$, $n_l=2.94$, $t=85$ nm, and $a=155$ nm. The GaAs/Al$_{x}$Ga$_{1-x}$As photonic crystal has the
following parameters: $n_h=3.57$, $n_l=3.125$, $t=80$ nm, and $a=150$ nm. (See Fig. \ref{fig:structure} for definition of parameters.) The band gap edges for
$D\rightarrow\infty$ are positioned at $a/\lambda$ equal to 0.1445 and 0.1634 for the GaAs/AlAs photonic crystal, and at $a/\lambda$ equal to 0.1431 and 0.1565
for the GaAs/Al$_{x}$Ga$_{1-x}$As photonic crystal. } \label{fig:bands}
\end{center}
\end{figure}

The rule of thumb generally used for designing microposts is to
make mirror layers one-quarter wavelength thick, and to choose the
optical thickness of the spacer equal to the target wavelength. In
the case of a planar DBR cavity (with $D\rightarrow\infty$), this
choice of parameters leads to the maximum reflectivities of the
mirrors and the maximum $Q$ factor of the cavity mode: the cavity
operates at the Bragg wavelength, for which the partial
reflections from all high- and low-refractive-index interfaces add
up exactly in phase. However, the strength of the cavity QED
phenomena is proportional to the ratio of the cavity $Q$ factor to
the mode volume $V$, as discussed in the previous section, and we
will try to design microposts in such a way that this ratio is
maximized.

In our earlier work \cite{ref:Pelton02}, we analyzed the $Q$
factor of the $HE_{11}$ mode in a GaAs/AlAs micropost as the
cavity diameter was tuned between 0.5 $\mu$m and 2 $\mu$m. The
remaining cavity parameters were chosen according to the
large-cavity rule of thumb, {\it i.e.} in such a way that the
cavity would operate at the Bragg wavelength for
$D\rightarrow\infty$. When the cavity diameter was decreased from
2 $\mu$m to 0.5 $\mu$m, the mode volume decreased by a factor of
almost ten, from $19.2(\lambda/n_h)^3$ to $2(\lambda/n_h)^3$,
while the cavity $Q$ dropped by only a factor of two, from $11
500$ to $5000$. Thus, in order to maximize the ratio of the
quality factor $Q$ to the mode volume $V$, we need to explore
structures with small diameters $D$, and try improve their $Q$
factors.

The reduction in $Q$ with decrease in $D$ is due to the
combination of two loss mechanisms: {\it longitudinal loss}
through DBR mirrors, and {\it transverse loss} due to imperfect
TIR confinement in the transverse direction. Let us address the
{\it longitudinal loss} first. The decrease in the post diameter
$D$ implies a change in the dispersion relation of the 1D photonic
crystal, and the size and position of its band gap, as illustrated
in Fig. \ref{fig:bands}. In this figure, it is assumed that the
high- and low-refractive-index regions of the photonic crystal
consist of GaAs and AlAs, with refractive indices of $n_h=3.57$
and $n_l=2.94$, and thicknesses of 70 nm and 85 nm, respectively,
or that they consist of GaAs and Al$_{x}$Ga$_{1-x}$As, with
refractive indices of $n_h=3.57$ and $n_l=3.125$, and thicknesses
of 70 nm and 80 nm, respectively. When the diameter $D$ decreases,
the frequencies of the band gap edges increase, and the size of
the band gap decreases. For structure diameters larger than 2
$\mu$m, band gap edges can be approximated by their values at
$D\rightarrow \infty$. Therefore, as $D$ decreases, the blue shift
of the cavity mode wavelength $\lambda$ increases relative to the
target wavelength at which the 1D cavity operates
\cite{ref:Pelton02}. Simultaneously, the size of the photonic
band-gap decreases, implying that the cavity mode is less confined
in the longitudinal direction than in the planar cavity case.

The cavity mode is strongly localized in real space, and
consequently delocalized in Fourier space ($k$-space), meaning
that it consists of a wide range of wave-vector components. Some
of these components are not confined in the post by TIR; {\em
i.e.}, they are positioned above the light line, where they can
couple to radiative modes, leading to {\it transverse loss}. A
cavity mode which is strongly confined in the longitudinal
direction by high-reflectivity mirrors is delocalized in Fourier
space and suffers large transverse loss. Similarly, a mode that is
delocalized in the longitudinal direction is more localized in
Fourier space and suffers less transverse loss. Therefore, when
optimizing the quality factor of three-dimensional microposts,
there is a tradeoff between these two loss mechanisms.

In the middle of a large band-gap, the longitudinal confinement is
strongest, but the $Q$ factor is limited by transverse loss. By
shifting the resonant wavelength away from the mid-gap ({\it
e.g.}, by tuning the thickness of the cavity spacer) one can
delocalize the mode in real space, localizing it more strongly in
Fourier space, reducing the contribution of wave-vector components
above the light line, and thereby decreasing the transverse
radiation loss. Eventually, as the mode wavelength approaches the
band-gap edges, the loss of longitudinal confinement starts to
dominate and $Q$ drops. Therefore, in the microposts with high
reflectivity mirrors and finite diameter, it is expected that the
maximum $Q$ will be located away from the mid-gap position.

\begin{figure}[hbtp]
\begin{center}
\subfigure{\epsfig{figure=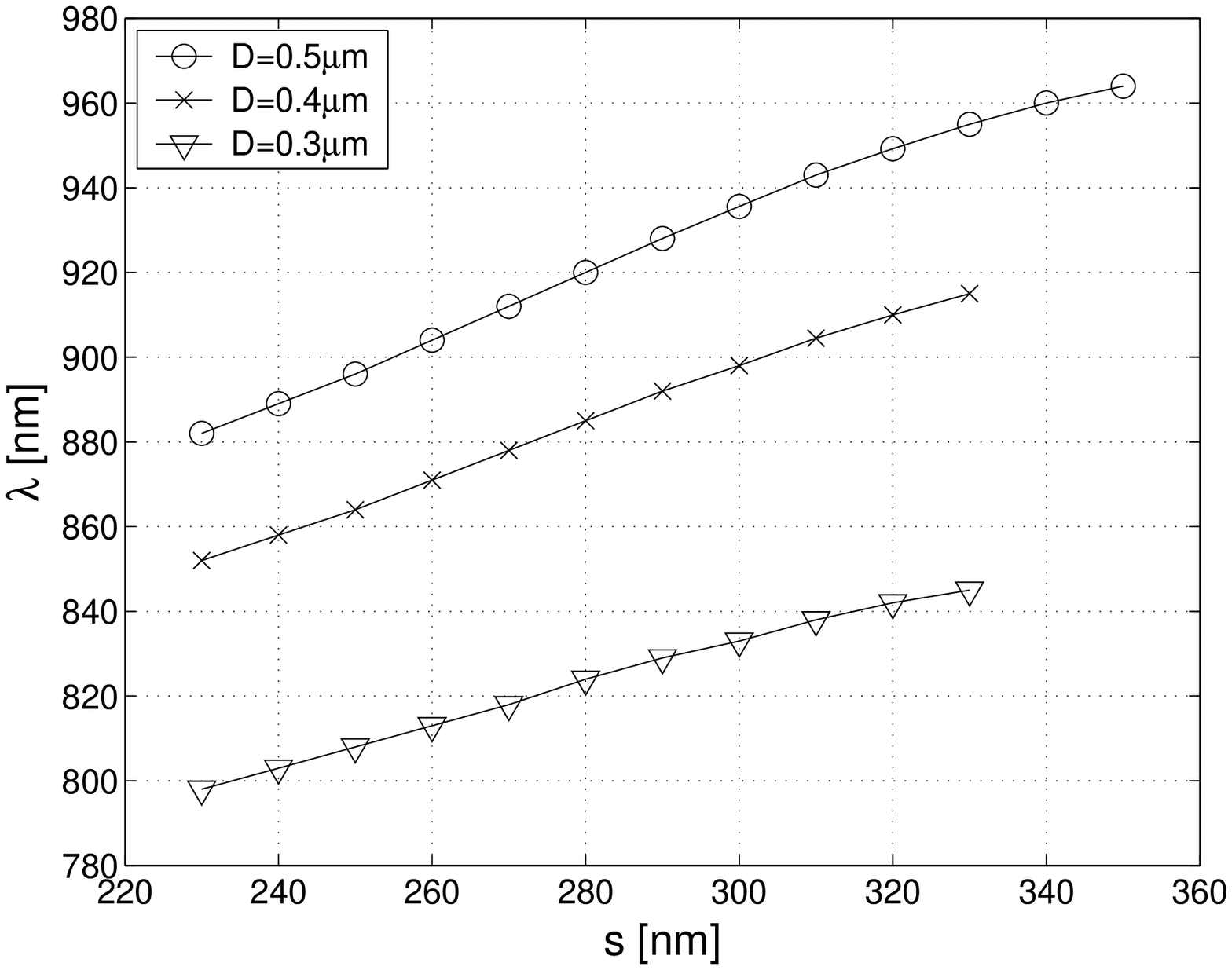,width=3in}}
\subfigure{\epsfig{figure=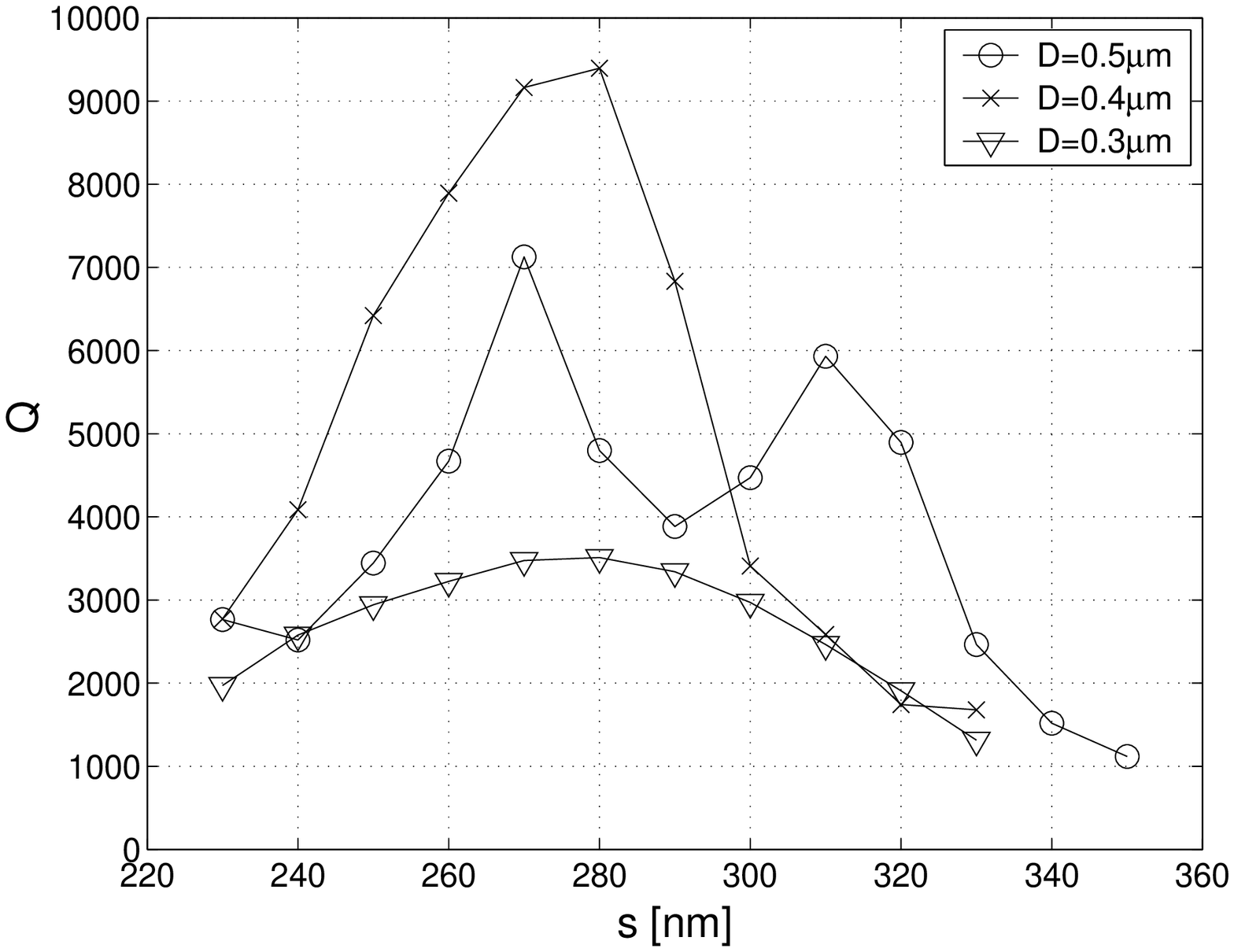,width=3in}}
\caption{Wavelength $\lambda$ and quality factor $Q$ of the fundamental mode in a
micropost with $a=155$ nm, $t=85$ nm, $n_h=3.57$, $n_l=2.94$, $MPT=15$ and $MPB=30$. The cavity diameter $D$ and the spacer thickness $s$ are tuned.}
\label{fig:vsrad1}
\end{center}
\end{figure}

Moreover, since the mode wavelength can be tuned from the mid-gap
towards any of the two band-gap edges, two local maxima of $Q$
({\em i.e.}, a double peak behavior in $Q$ {\em vs.} mode
wavelength) are expected. Besides detuning the mode wavelength
from the mid-gap, we can also suppress the transverse loss by
relaxing the mode slightly in the longitudinal direction, {\it
i.e.}, by reducing the reflectivities of photonic crystal mirrors
and decreasing the band gap size. This can be achieved by
shrinking the cavity diameter, or by changing the photonic crystal
parameters ({\it e.g.}, by reducing the refractive-index
contrast).

In this article, we study both of these approaches to $Q$
optimization: tuning the mode wavelength away from the mid-gap by
changing the spacer thickness, and tuning the mirror
reflectivities by changing photonic crystal parameters or cavity
diameter. We also show that the employment of very high
reflectivity mirrors cannot lead to high-$Q$ cavities with small
diameters, as the transverse radiation loss is high, resulting
from very strong mode localization in the longitudinal direction.

\section{Maximizing the ratio of quality factor to mode volume
for the fundamental mode in a micropost microcavity}

\subsection{Tuning the cavity diameter and the cavity spacer}

In our earlier work \cite{ref:Pelton02}, we analyzed the $Q$
factor of the $HE_{11}$ mode in a GaAs/AlAs micropost as the
cavity diameter was tuned between 0.5 $\mu$m and 2 $\mu$m. The
remaining cavity parameters were chosen in such a way that the
cavity would operate at the Bragg wavelength for
$D\rightarrow\infty$ ($a=155$ nm, $t=85$ nm, $s=280$ nm,
$n_h=3.57$ and $n_l=2.94$). The number of DBR mirror pairs on top
and bottom were $MPT=15$ and $MPB=30$, respectively.

Let us first study the $HE_{11}$ mode as the diameter is decreased
below 0.5 $\mu$m, keeping all other structure parameters the same
as above. In order to tune the mode frequency within the band gap,
we tune the spacer thickness $s$. Results for $\lambda$, $Q$, $V$,
and $Q/V$ are shown in Figs. \ref{fig:vsrad1} and
\ref{fig:vsrad2}. From Fig. \ref{fig:bands}, we see that the band
gaps in these structures extend from 875 nm to 969 nm, from 850 nm
to 920 nm, and from 790 nm to 850 nm, for structure diameters of
0.5 $\mu$m, 0.4 $\mu$m, and 0.3 $\mu$m, respectively. As we have
noted previously, when $D$ decreases, the band gap edges shift
towards lower wavelengths, and the size of the band gap decreases.
The cavity mode wavelength is blue-shifted in this process, as can
be seen in Figure \ref{fig:vsrad1}.

The mode volume $V$ is minimized when the mode wavelength is
located near the middle of the band gap. For the structures with
$D$ equal to 0.4 $\mu$m and 0.3 $\mu$m, the maximum $Q$ factor
also occurs close to the mid-gap. Different behavior is seen for
the structure with $D$ equal to 0.5 $\mu$m, which has a local
minimum of $Q$ at mid-gap and exhibits a double-peak behavior.

The double-peak behavior was already introduced in the previous
section. In the middle of the band gap, where the longitudinal
mode confinement is strongest and the mode volume is minimum, the
radiation loss in the transverse direction is high, and the $Q$
factor is degraded. By shifting the resonant wavelength away from
the mid-gap, the mode is delocalized in real space, leading to a
reduction in the transverse radiation loss ({\em e.g.}, at the
positions of the two peaks in $Q$). Eventually, as the mode
wavelength approaches the band-gap edges, the loss of longitudinal
confinement starts to dominate, $Q$ drops, and the mode volume
increases.

\begin{figure}[hbtp]
\begin{center}
\subfigure{\epsfig{figure=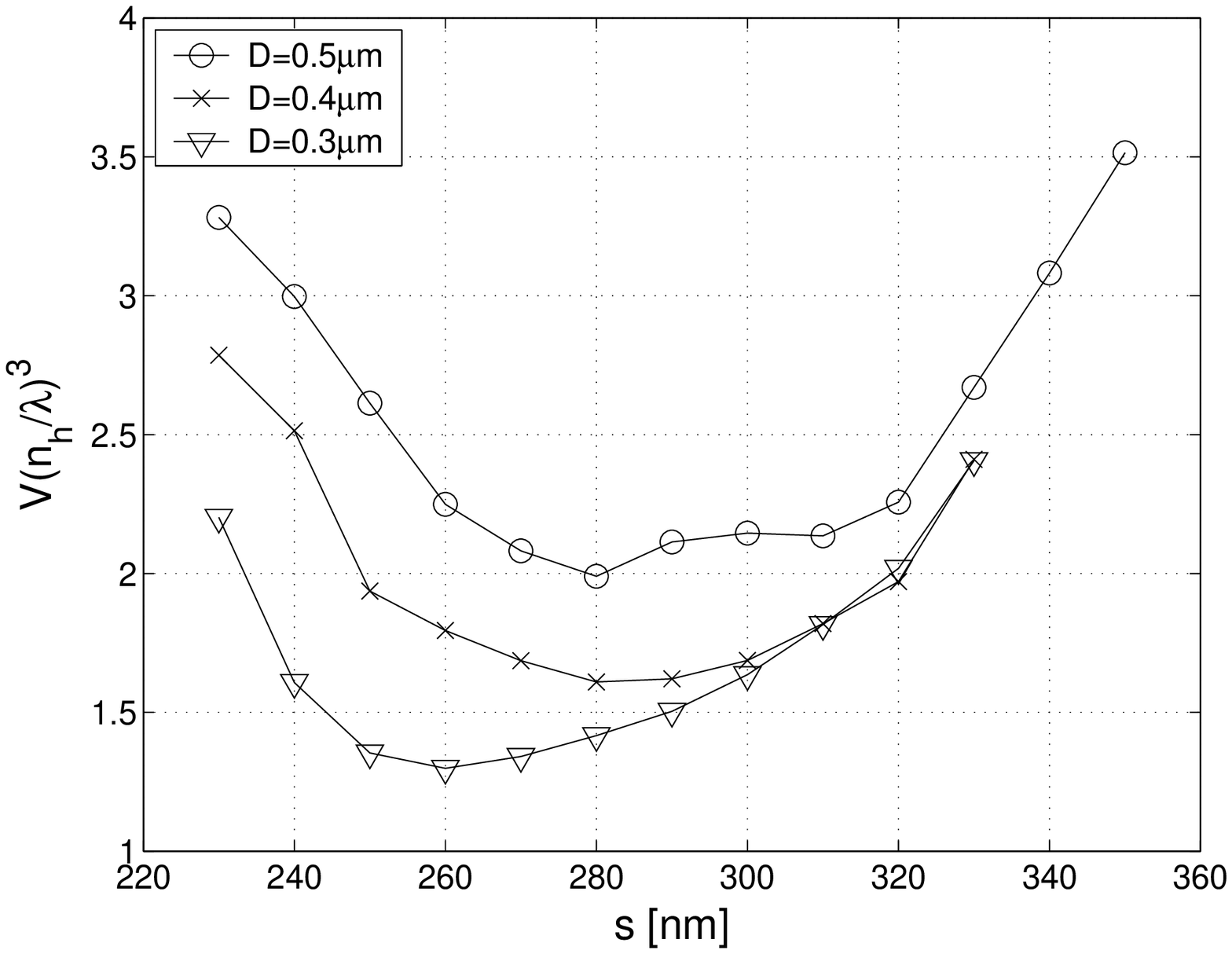,width=3in}}
\subfigure{\epsfig{figure=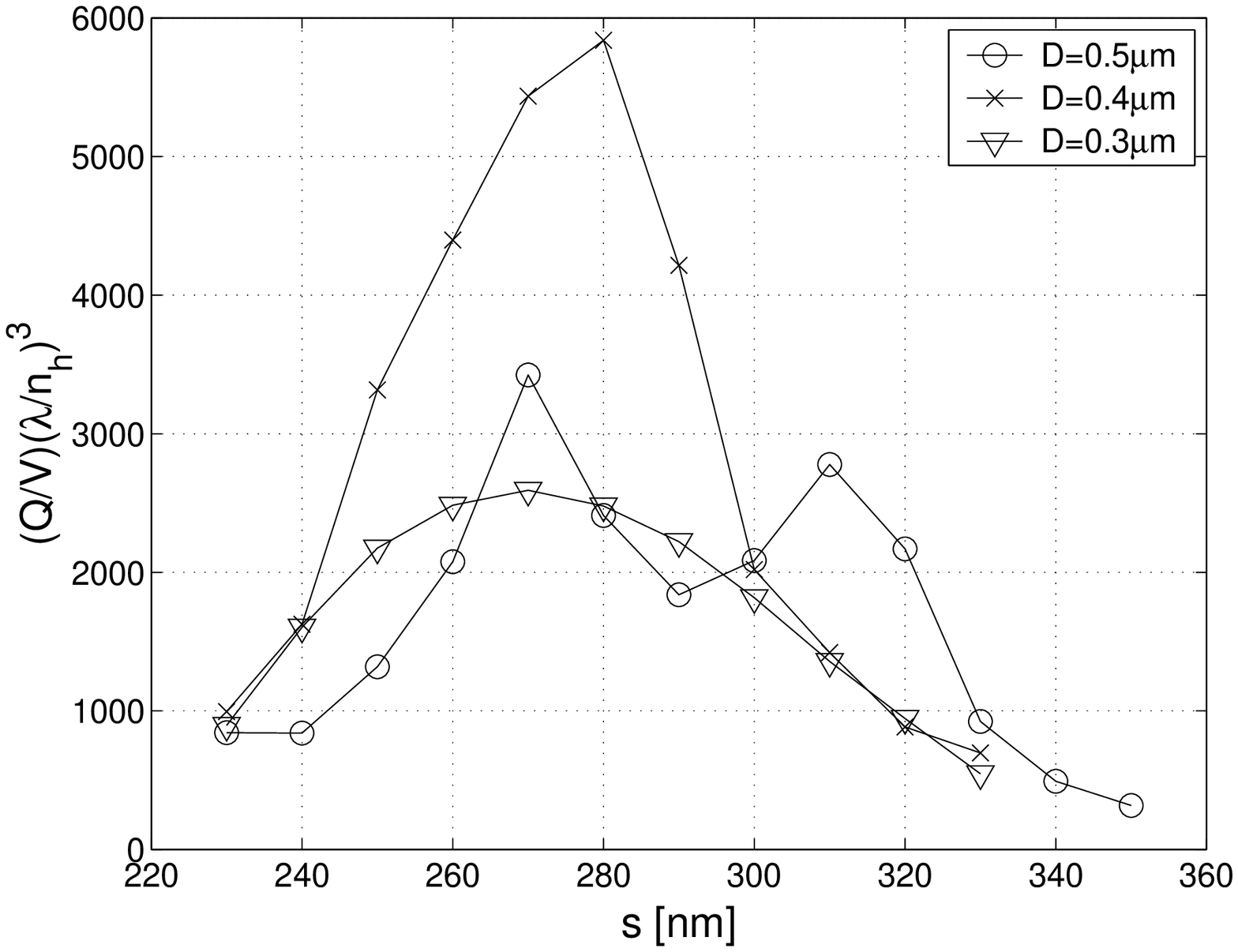,width=3in}}
\caption{Mode volume $V$ and ratio of quality factor $Q$ to $V$ for the $HE_{11}$ mode in a micropost with $a=155$ nm, $t=85$ nm, $n_h=3.57$, $n_l=2.94$,
$MPT=15$ and $MPB=30$. The cavity diameter $D$ and the spacer thickness $s$ are tuned.} \label{fig:vsrad2}
\end{center}
\end{figure}

To support this explanation, we analyze the same structure, with
$D=0.5$ $\mu$m, but with the number of mirror pairs on top ($MPT$)
increased from $15$ to $25$. As expected, at mid-gap, $Q$ does not
increase significantly with $MPT$. The mode there is already
strongly confined in the longitudinal direction, and the addition
of extra pairs does not change the longitudinal loss. The modal
$Q$ factor is determined by the radiation loss in the transverse
direction, which is independent of $MPT$. On the other hand, the
$Q$'s at the two peaks increase with $MPT$. At these points, the
mode is not confined as well in the longitudinal direction, and
longitudinal loss can be reduced by adding more mirror pairs.

As an even stronger demonstration of our explanation for the
double-peak behavior, we separate the radiation loss into the loss
above the top micropost surface ($L_a$), and the loss below it
($L_b$). The total $Q$ is a combination of two newly introduced
quality factors, $Q_a$ and $Q_b$, which are inversely proportional
to $L_a$ and $L_b$, respectively:
\begin{equation}
\frac{1}{Q}=\frac{1}{Q_a}+\frac{1}{Q_b} \, .
\end{equation}
It follows from their definition that $Q_a$ and $Q_b$ are measures
of the longitudinal and transverse loss, respectively. We analyze
two sets of structure parameters, corresponding to the local
maximum or minimum in $Q$. For $s=270$ nm and $D=0.5$ $\mu$m
(local maximum), we calculate $Q_a \approx 14 500$ and $Q_b\approx
13 910$, while, for $s=290$ nm and $D=0.5$ $\mu$m (local minimum),
we calculate $Q_a\approx 16 000$ and $Q_b\approx 5100$. These
results show that the local minimum in $Q$ is due to an increase
in the transverse loss, manifested as a drop in $Q_b$.

\begin{figure}[hbtp]
\begin{center}
\epsfig{figure=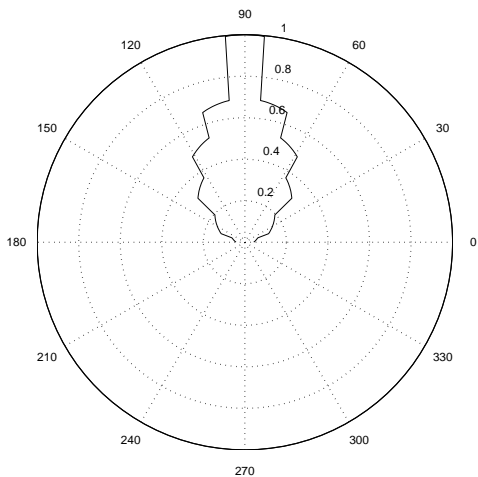,width=2.5in}
\caption{Radiation pattern from the $HE_{11}$ mode in a micropost with the following parameters: $a=155$ nm,
$t=85$ nm, $D=0.5$ $\mu$m, $s=270$ nm, $n_h=3.57$, $n_l=2.94$, $MPT=15$ and $MPB=30$. An angle of $90^o$ corresponds to the vertical axis of the micropost.}
\label{fig:radpat}
\end{center}
\end{figure}

Let us now address the single-peak behavior of $Q$ as a function
of cavity spacer thickness, when $D$ is equal to 0.4 $\mu$m or 0.3
$\mu$m. Structures with smaller diameters have smaller band gaps,
as illustrated in Fig. \ref{fig:bands}, and the cavity modes are
more delocalized in the longitudinal direction, relative to the
structure with $D=0.5$ $\mu$m. The defect modes must therefore be
more localized in Fourier space, and will thus suffer less
radiation loss in the transverse direction. This implies that the
$Q$ factors are determined mostly by the longitudinal loss.  They
reach their maxima at the mid-gap, where the mode volume is
minimum, and the longitudinal confinement is strongest.

The maximum $Q/V$ ratio of almost $6000$ (where $V$ is measured in
cubic wavelengths in the high-refractive index material) is
achieved for the structure with $D=0.4$ $\mu$m. For this
structure, the $Q$ factor is close to $9500$, and the mode volume
is $1.6(\lambda/n_h)^3$. For $D=0.4$ $\mu$m, a variation in the
thicknesses of the mirror layers allows us to achieve a small
increase in the $Q$ factor, to $10 500$, and in the $Q/V$ ratio,
to $6500$. This result is obtained for $a=155$ nm, $t=75$ nm, and
$s=290$ nm.

\begin{figure}[hbtp]
\begin{center}
\subfigure{\epsfig{figure=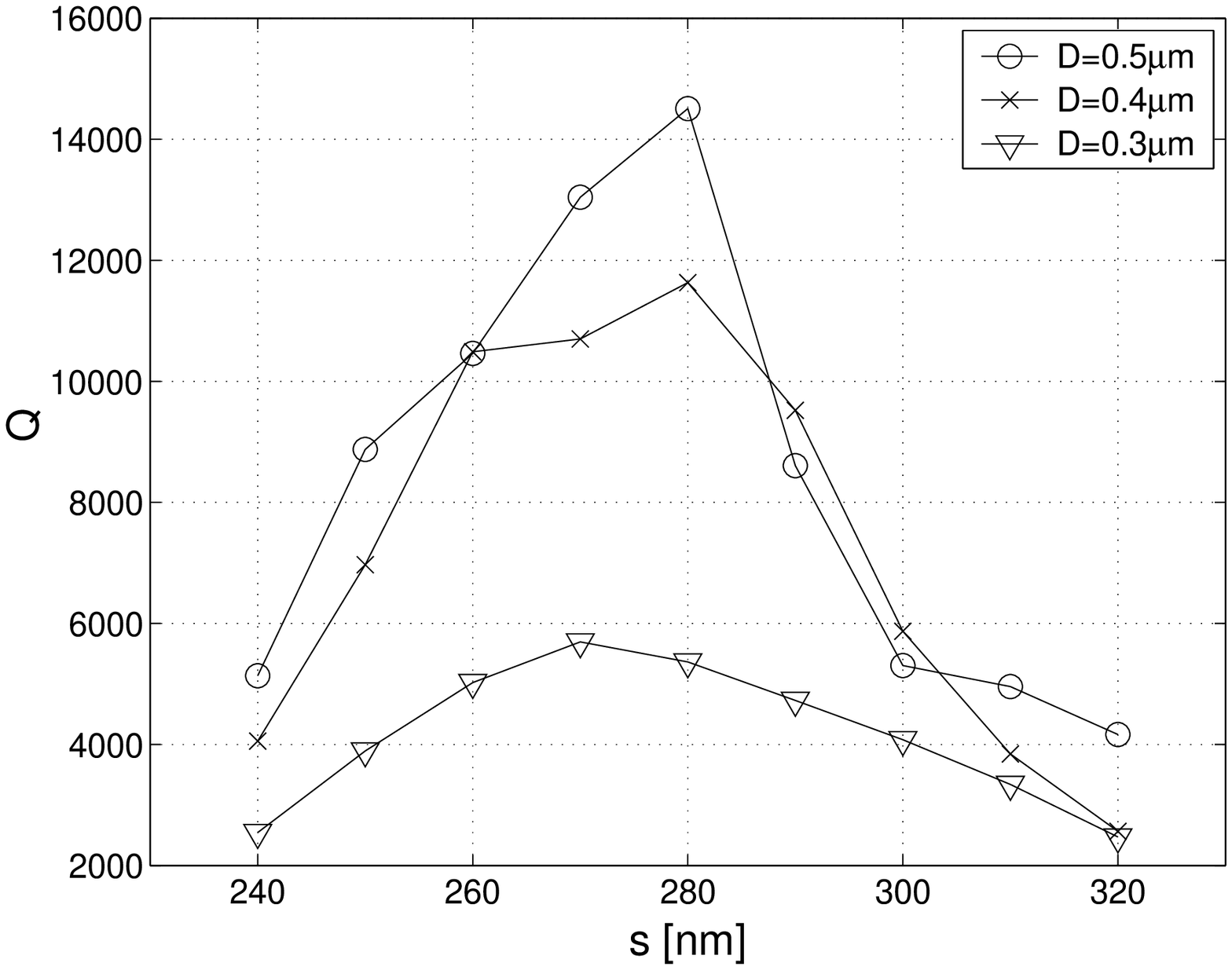,width=3in}}
\subfigure{\epsfig{figure=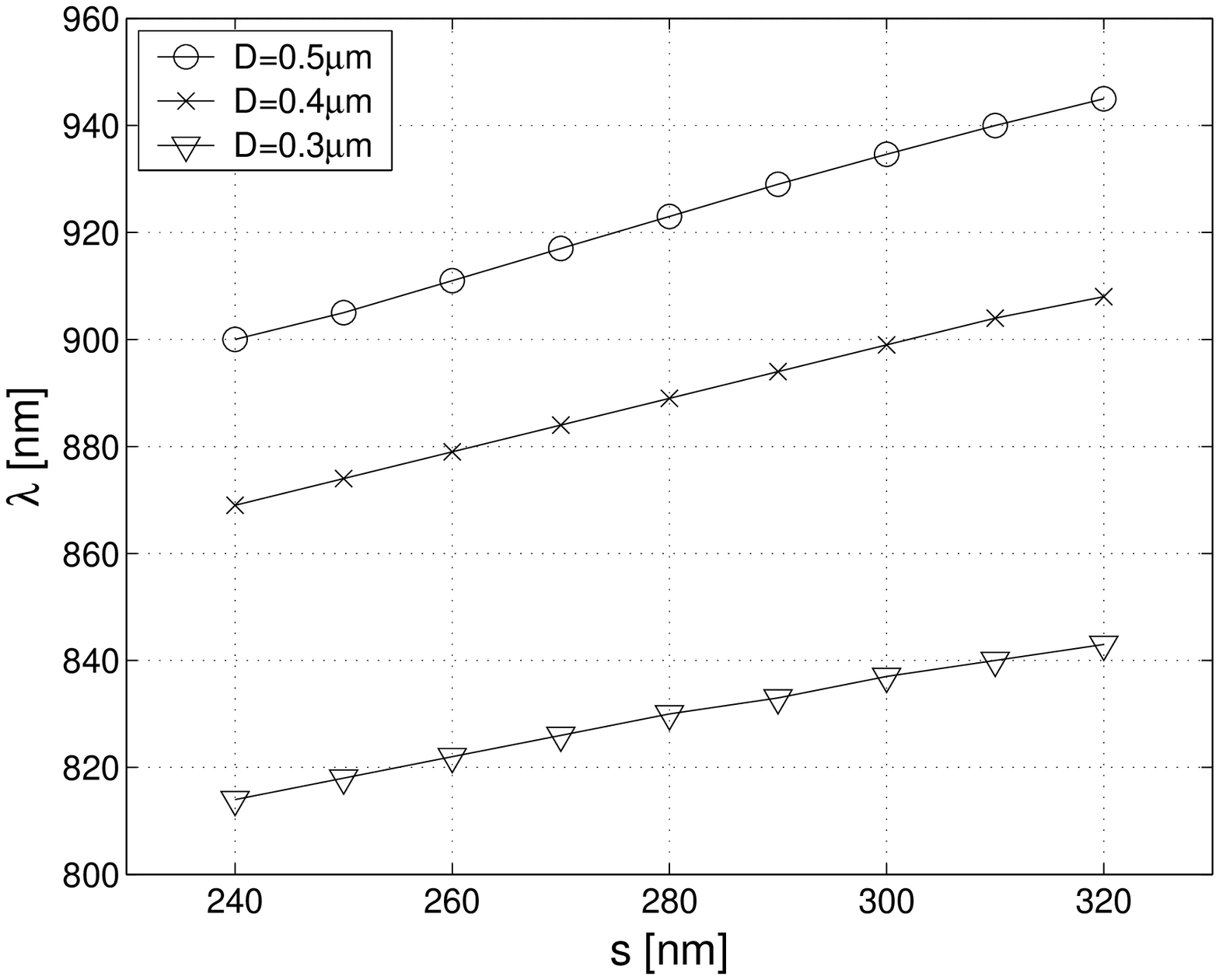,width=3in}}
\caption{Quality factor $Q$ and wavelength $\lambda$ of the $HE_{11}$ mode
in a micropost with $a=150$ nm, $t=80$ nm, $n_h=3.57$, $n_l=3.125$, $MPT=25$ and $MPB=30$. The cavity diameter $D$ and the spacer thickness $s$ are tuned.}
\label{fig:algaas_vs_rad1}
\end{center}
\end{figure}

In the introduction, we mentioned that an advantage of microposts,
relative to other solid state microcavities, is that the light
escapes from them in a single-lobed Gaussian-like pattern, normal
to the sample surface. In order to show this, we calculate the
far-field radiation pattern from a micropost with $D=0.5$ $\mu$m
and $s=270$ nm. We are unable to directly compute the far field by
employing the FDTD method, as we are limited by our computer
memory size. However, we can estimate the far field from the
Fourier transform of the near field, using the method described in
Ref. \cite{ref:JV2002}. The calculated radiation pattern is shown
in Fig. \ref{fig:radpat}. Its resolution is limited by the
resolution that we can achieve in Fourier space, or, more
precisely, by the number of pixels in the light cone.  This, in
turn, is dictated by the size of the computational domain. The
best resolution in Fourier space that we can obtain with a
reasonable size of the computational domain is seven pixels per
light cone radius. Nonetheless, the computed radiation pattern
demonstrates that even microposts with small diameters can emit
light in a Gaussian-like pattern. The FWHM of the emission lobe
shown is approximately equal to $50^o$.

\subsection{Other material systems}

\subsubsection{GaAs/Al$_x$Ga$_{1-x}$As cavities}

In the previous section of this article, we stated that a
potential route to maximizing $Q$ for small micropost diameters is
the construction of a photonic crystal with a small
refractive-index perturbation.  As the perturbation gets smaller,
the cavity mode becomes more delocalized in real space, and
consequently more localized in Fourier space.  This, in turns,
leads to reduction in the transverse radiation loss. Furthermore,
the cavity resonance can be located at lower frequencies, where
the density of free-space radiation modes is smaller. In order to
compensate for the increased longitudinal loss, we need to put
more mirror pairs on top of these structures.

We will now analyze a micropost with the following parameters:
$a=150$ nm, $t=80$ nm, $MPT=25$, $MPB=30$, $n_h=3.57$, and
$n_l=3.125$. This choice of refractive indices corresponds to
GaAs/Al$_{x}$Ga$_{1-x}$As layers. Both the cavity diameter $D$ and
the spacer thickness $s$ are tuned. The positions of the band gap
edges as a function of $D$ are illustrated in Fig.
\ref{fig:bands}. By comparing to the positions of the band gap
edges for the GaAs/AlAs system, we confirm that the band gap in
the GaAs/Al$_x$Ga$_{1-x}$As system is shifted to lower
frequencies, and that its size is decreased. This affects the
$HE_{11}$ mode dramatically, as can be seen in Figs.
\ref{fig:algaas_vs_rad1} and \ref{fig:algaas_vs_rad2}.

By comparing Fig. \ref{fig:algaas_vs_rad2} to Fig.
\ref{fig:vsrad2}, we can see that the mode volume increases when
the refractive-index contrast is reduced, as a result of the
reduction in band-gap size. Even though $Q$ larger than 14000 can
be achieved for $D=0.5$ $\mu$m, $V$ also increases, and the
maximum $Q/V$ ratio is similar to that calculated for the
GaAs/AlAs system. Furthermore, this $Q/V$ ratio can be achieved in
the GaAs/AlAs system with fewer top mirror pairs. Longitudinal
loss dominates in the GaAs/Al$_x$Ga$_{1-x}$As system, and $Q$ {\em
vs.} $s$ plots demonstrate a single-peak behavior.

When the number of mirror pairs on top is reduced from 25 to 20,
the peak $Q$ factor of the GaAs/Al$_x$Ga$_{1-x}$As micropost with
diameter of 0.5 $\mu$m drops from around 14000 to 4000, showing
that the the longitudinal loss is dominant in this case, and a
large number of mirror pairs
is necessary to achieve large $Q$ factors.\\

\begin{figure}[t]
\begin{center}
\subfigure{\epsfig{figure=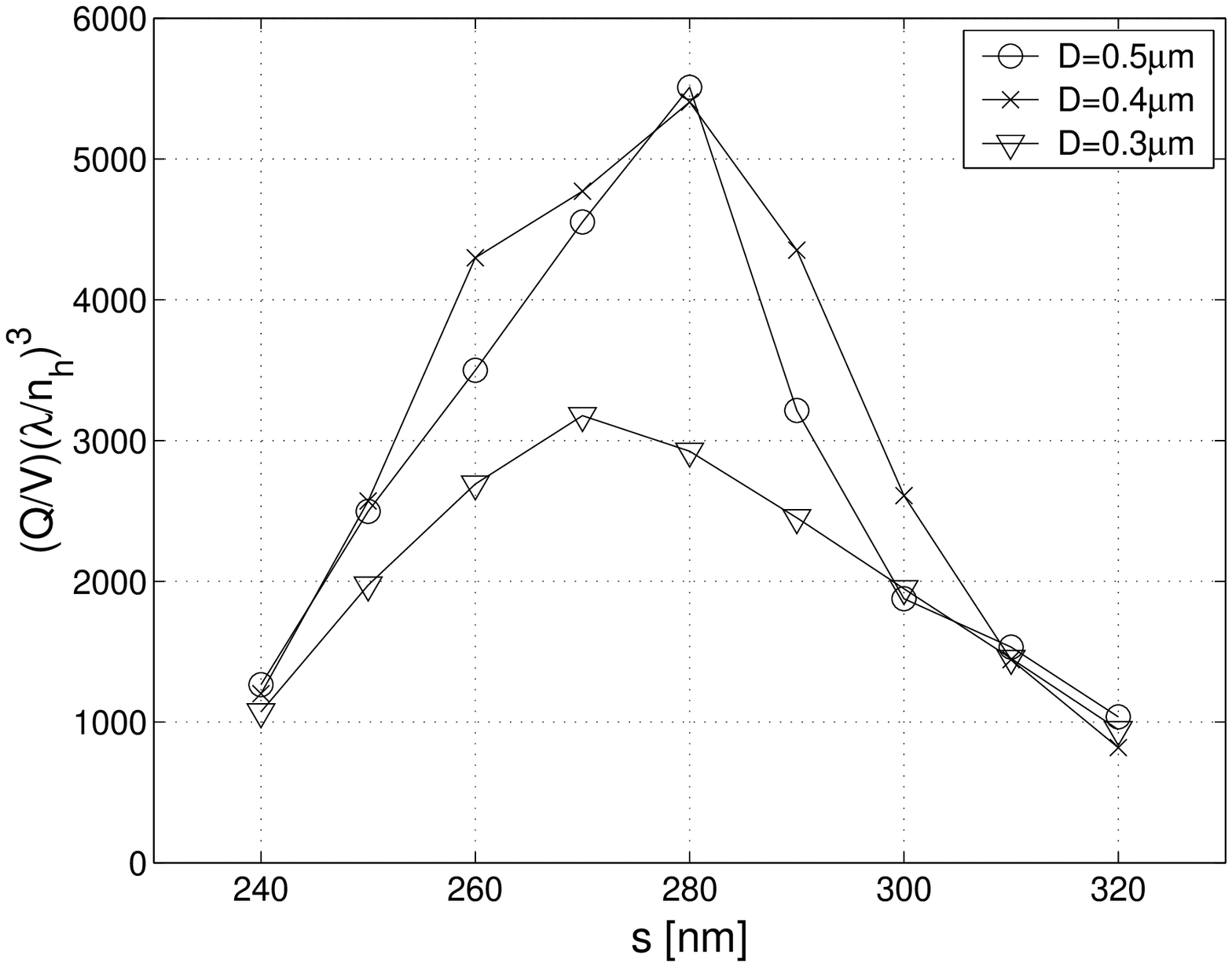,width=3in}}
\subfigure{\epsfig{figure=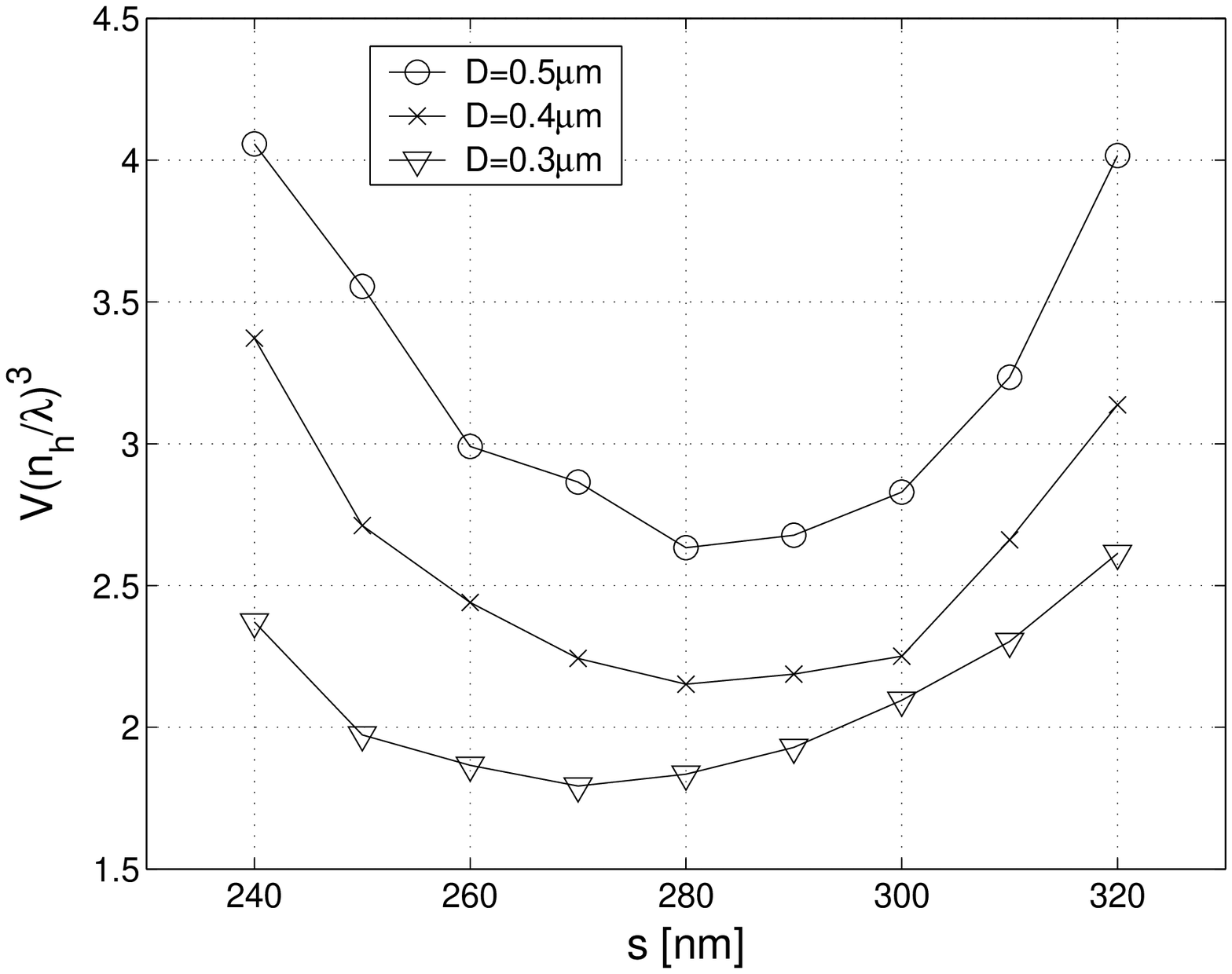,width=3in}}
\caption{Ratio of quality factor $Q$ to mode volume $V$, and mode volume $V$
for the $HE_{11}$ mode in a micropost with $a=150$ nm, $t=80$ nm, $n_h=3.57$, $n_l=3.125$, $MPT=25$ and $MPB=30$. The cavity diameter $D$ and the spacer
thickness $s$ are tuned.} \label{fig:algaas_vs_rad2}
\end{center}
\end{figure}

\subsubsection{GaAs/AlO$_x$ cavities}

From the results already presented in this article, it is clear
that a material system with a high refractive-index contrast, such
as GaAs/AlO$_x$, is not a good choice for high $Q$, small
mode-volume microposts. High refractive index contrast can
certainly produce larger band gaps, and thereby provide a better
longitudinal confinement of the cavity mode. However, if the
contrast is increased, the mode suffers more radiation loss in the
transverse direction, which limits its $Q$ factor. To confirm
this, we analyzed a structure with $n_h=3.57$, $n_l=1.515$,
$a=235$ nm, $t=165$ nm, $MPT=15$, and $MPB=30$, for different $D$.
We were unable to obtain good mode localization for $D<0.8$
$\mu$m, and the calculated $Q$ factors were under $250$. For
$D=0.8$ $\mu$m, the mode has $Q=600$ and $\lambda=947$ nm. If we
keep increasing $D$ to 1.3 $\mu$m, $Q$ factors remain below
$1000$.

\section{Cavity quantum electrodynamics with microposts}

The question that we would like to address in this section is
whether such cavity QED phenomena as strong coupling or single-dot
lasing can be observed in the optimized microposts. Let us revisit
our best design, with $Q\approx 10^4$, $V=1.6(\lambda/n_h)^3$,
$D=0.4 \mu$m, $\lambda=885$ nm, and the cavity field decay rate
$\kappa=(\pi c)/(\lambda Q)=106$ GHz.

By combining Eqns. \ref{eqn:g} and \ref{eqn:gamma0}, the Rabi
frequency $g_0$ of a system on resonance can be expressed as
\begin{equation}
g_0=\frac{\Gamma_0}{2}\sqrt{\frac{V_0}{V}}, \label{eqn:g1}
\end{equation}
where $V_0=(3c\lambda^2\epsilon_0)/(2\pi\Gamma_0\epsilon_M)$. Let
us assume that a quantum dot exciton without a cavity has a
typical homogenous linewidth $\gamma_h=20$ GHz, and a radiative
lifetime of 0.5 ns, corresponding to a spontaneous emission rate
of $\Gamma=2$ GHz. The free-space spontaneous emission rate is
$\Gamma_0=\Gamma/n_h=0.56$ GHz. The Rabi frequency for our
optimized cavity, calculated from Eqn. \ref{eqn:g1}, is equal to
$g_0=400\Gamma_0=224$ GHz. If we assume that the quantum dot is
located in the center of the micropost and that its dipole is
aligned with the electric field, we have $g=g_0$. Strong coupling
is therefore possible in this case, since
$g_0>\kappa,\pi\gamma_h$. The minimum quality factor necessary to
achieve strong coupling is approximately equal to 5000. This
provides a reasonable margin for $Q$ degradation due to
fabrication imperfections.

Is strong coupling possible with larger diameter microposts, such
as $D=2 \, \mu$m? The mode volume in such a structure is on the
order of $20(\lambda/n_h)^3$, as we mentioned previously. For the
same quantum dot, with $\Gamma_0=0.56$ GHz, placed in the center
of this large cavity, the Rabi frequency is $g_0=60$ GHz. For our
experimentally observed homogenous broadening $\gamma_h=20$ GHz,
it is impossible to reach strong coupling, since
$\pi\gamma_h>g_0$. Even if the homogenous linewidth were reduced
to 2 GHz ({\em i.e.}, if the homogeneous broadening were entirely
due to radiative decay), the $Q$ factor required to achieve strong
coupling would be on the order of $2\times10^4$. We therefore
conclude that large-diameter microposts are not promising
candidates for the observation of strong coupling with a single
quantum dot.

Designs of two-dimensional photonic crystal microcavities in
free-standing membranes were recently proposed that allow for very
strong coupling between the cavity field and a neutral atom
trapped in one of photonic crystal holes \cite{ref:JV2001}. We
will now address the feasibility of strong coupling with a single
quantum dot in these structures, and compare them to our micropost
designs. These photonic-crystal microcavities can localize light
into mode volumes equal to $1/2(\lambda/n_h)^3$, with $Q$ on the
order of $10^4$. However, since the field intensity is strongest
in or around the defect air hole (where a neutral atom would be
trapped), it is almost impossible to place a quantum dot at the
point where its interaction with the cavity field would be
strongest. For example, if the dot is placed at the point where
the field intensity is $60\%$ of its maximum value, the Rabi
frequency remains the same as for our best micropost design
($g=224$ GHz), despite a three-fold decrease in the mode volume.
The quality factor is in the same range as for the optimized
microposts, which implies that the potential of these structures
to achieve strong coupling with single quantum dots is similar to
that of microposts.

What about single-dot lasing in microposts? The lasing condition
for such a microscopic system is given by Eqn.
\ref{eqn:threshold}. Clearly, in order to reach the laser
threshold, it is necessary to increase the Purcell factor $F$ and
the quality factor $Q$. Our analysis indicates that large
spontaneous-emission enhancement is possible in microposts. As an
example, let us consider an unoptimized microcavity with the
following parameters: $n_h=3.57$, $n_l=2.94$, $D=0.5$ $\mu$m,
$s=280$ nm, $a=155$ nm, $t=85$ nm, $MPT=15$, $MPB=30$, $Q=4800$,
$\lambda=920$ nm and $V=2(\lambda/n_h)^3$. (The method used for
calculation is described in Refs. \cite{ref:JV99} and
\cite{ref:Xu99}.) The Purcell factor for an emitter with zero
linewidth positioned in the center of this micropost, is equal to
147.  The enhancement drops to 65 for a linewidth of 100 GHz. Such
a high Purcell factor would imply that $\beta\approx 1$. Let us
also assume that $N_A \approx 1$, corresponding to fast pumping.
In order to observe single-dot lasing, we then need to satisfy the
condition $\tau_{ph}>\tau_{sp}$. For a cavity with $Q=10^4$
operating at $\lambda\approx1$ $\mu$m, we have
$\tau_{ph}\approx5.3$ ps. Therefore, to achieve single-dot lasing,
we would need $\tau_{sp}$ shorter than $5$ ps. If we again assume
that the lifetime of an exciton without a cavity is $0.5$ ns,
corresponding to a spontaneous emission rate of 2 GHz, lifetime
reduction to $5$ ps would require a Purcell factor equal to 100.
As mentioned above, such Purcell factors are possible for
sufficiently narrow homogeneous linewidths. Single-dot lasing
should therefore be possible in the optimized microposts.

\section{Conclusions}

Using the FDTD method, we have analyzed the fundamental
($HE_{11}$) mode in ideal, three-dimensional micropost cavities,
for a variety of material systems (GaAs/AlAs,
GaAs/Al$_x$Ga$_{1-x}$As and GaAs/AlO$_x$). Microcavities were
treated as single defects in a 1D cylindrical photonic crystal,
which allowed us to push the limits of quality factors and mode
volumes $V$ beyond those achievable by standard micropost designs.
Our motivation was to maximize the $Q/V$ ratio of the defect mode,
in order to use cavity-QED phenomena to build  novel
optoelectronic devices, such as single-dot lasers and
high-efficiency light-emitting diodes, or to construct hardware
for quantum computers and quantum communication systems, such as
single-photon sources and strongly coupled quantum dot -- cavity
systems.

The standard approach for designing micropost microcavities is to
choose the thicknesses of mirror layers and the spacer
corresponding to the Bragg wavelength of a planar microcavity. We
have shown that this approach does not necessarily lead to the
highest $Q$ factors for the small cavity diameters analyzed in
this article ($D\leq0.5$ $\mu$m). Another widespread misconception
is that the $Q$ of the cavity mode can always be improved by
increasing the refractive-index contrast of the mirror layers. We
have shown that this approach fails for small post diameters. Two
primary loss mechanisms in three-dimensional microposts are the
loss in the longitudinal direction, through DBR mirrors, and the
loss in the transverse direction, due to imperfect confinement by
TIR. A cavity mode which is strongly confined in the longitudinal
direction by high-reflectivity mirrors is delocalized in Fourier
space, leading to increased coupling to radiation modes and
increased transverse loss. Similarly, a mode that is delocalized
in the longitudinal direction and suffers significant longitudinal
loss is more localized in Fourier space and suffers less
transverse loss. When designing three-dimensional microposts,
there is a tradeoff between these two loss mechanisms.

We were able to achieve $Q$ as high as $10^4$ together with mode
volume as small as $1.6(\lambda/n_h)^3$ by optimizing structure
parameters. Even though this range of values can be achieved in
both GaAs/AlAs and GaAs/Al$_x$Ga$_{1-x}$As material systems, the
former is a better choice from the perspective of fabrication,
since the optimized structures require fewer mirror pairs on top.

We have also demonstrated that the optimized cavities can be used
to observe novel cavity-QED phenomena, such as single-dot lasing
or strong coupling between a single quantum dot and the cavity
field. Moreover, the potential of microposts to achieve strong
coupling with quantum dots is comparable to that of the largest
$Q/V$ planar photonic crystal microcavities that are presently
known \cite{ref:JV2001}.

\begin{bf}Acknowledgement \end{bf}\\\\
This work was supported in part by the ARO, under the research grant 0160-G-BC575 (MURI program entitled ``Single photon turnstile devices"). Financial
assistance for M.P. was provided by the Stanford Graduate Fellowships program.\\
Present address of M. Pelton: Laboratory of Quantum Optics and Quantum Electronics,
 Department of Microelectronics and Information Technology, Royal Institute of Technology (KTH),
 Electrum 229, SE-164 40  Kista, Sweden.\\
 Address if A. Scherer: California Institute of Technology, Pasadena, CA 91125.

\bibliographystyle{unsrt}
\bibliography{YY2_ref_rev2}

\end{document}